\documentstyle[prd,aps,floats]{revtex}

\def\ds{\displaystyle}

\begin{document}

\draft
\input epsf
\twocolumn[\hsize\textwidth\columnwidth\hsize\csname
@twocolumnfalse\endcsname

\title{Q-ball formation in the gravity-mediated SUSY breaking scenario}

\author{S. Kasuya$^{1}$ and M. Kawasaki$^{2}$}
\address{${}^1$ Department of Physics, Ochanomizu University,
  Bunkyo-ku, Tokyo 112-8610, Japan}
\address{${}^2$ Research Center for the Early Universe, University 
  of Tokyo, Bunkyo-ku, Tokyo 113-0033, Japan}

\date{February 28, 2000}

\maketitle

\begin{abstract}
We study the formation of Q-balls which are made of flat
directions that appear in the supersymmetric extension of the standard
model in the context of gravity-mediated supersymmetry breaking.
The full non-linear calculations for the dynamics of the complex
scalar field are made. Since the scalar potential in this model is
flatter than $\phi^2$, we have found that fluctuations develop and go
non-linear to form non-topological solitons, Q-balls. The size of a
Q-ball is determined by the most amplified mode, which is completely
determined by the model parameters. On the other hand, the charge of
Q-balls depends linearly on the initial charge density of the 
Affleck-Dine (AD) field. Almost all the charges are absorbed into
Q-balls, and only a tiny fraction of the charges is carried by a relic 
AD field. It may lead to some constraints on the baryogenesis and/or
parameters in the particle theory. The peculiarity of
gravity-mediation is the moving Q-balls. This results in collisions
between Q-balls. It may increase the charge of Q-balls, and change its
fate. 
\end{abstract}

\pacs{PACS numbers: 98.80.Cq, 11.27.+d, 11.30.Fs
      \hspace{5cm} hep-ph/0002285}


\vskip2pc]

\setcounter{footnote}{1}
\renewcommand{\thefootnote}{\fnsymbol{footnote}}

\section{Introduction}
A Q-ball is a kind of a non-topological soliton, whose stability is
guaranteed by some conserved charge in scalar field theory
\cite{Coleman,Kusenko1}. It can be made of the scalar fields which
appear as flat directions in the supersymmetric extension of the
standard model \cite{Kusenko2,Dvali}. Particularly, in the
minimal supersymmetric standard model (MSSM), the baryon and/or lepton
number are the conserved charges, since those flat directions consist
of squarks and/or sleptons \cite{Dine}. It is known that large Q-ball
solutions exist when both gauge-mediated and gravity-mediated
supersymmetry (SUSY) breaking scenarios are included
\cite{KuSh,EnMc1}. In the gauge-mediation scenario, the baryonic
charged Q-ball, the B-ball, is stable against decay into nucleons,
since the energy per unit charge becomes less than the nucleon mass, 1
GeV, for large enough Q-ball charge:  $E\sim mQ^{3/4}$ \cite{KuSh}. 
Therefore, large B-balls can be a promising candidate for the cold
dark matter. On the other hand, Q-ball energy grows linearly in the
gravity-mediation scenario: $E\sim mQ$ \cite{EnMc2}. They can thus
decay into both nucleons (baryons) and lightest supersymmetric
particles (LSPs), which become the dark matter in the universe. In the
both scenarios, we can expect a close relation between the energy
density of the baryon and dark matter such as 
$\Omega_b \sim \Omega_{DM}$ \cite{KuSh,EnMc2} ($\Omega_b$ and
$\Omega_{DM}$ are density parameters of the baryon and the dark
matter, respectively). In particular, a somewhat more definite
relation on the 
number densities hold for the gravity-mediation scenario: 
$n_{LSP} \simeq N_B f_B n_b$ \cite{EnMc2,EnMc3}. Here $N_B$ is the
number of LSP decay products from the scalar field (flat direction) 
with unit baryon number, and $f_B$ is the fraction of the charge
stored in the form of Q-balls. For these mechanism to work, the charge
of B-ball should be in the range $10^{20}-10^{30}$ \cite{LaSh,EnMc2}.

Those large Q-balls are expected to be created through Affleck-Dine
(AD) mechanism \cite{AfDi} in the inflationary universe
\cite{KuSh,EnMc1,EnMc2}. The coherent state of the AD scalar field
which consists of some flat direction in MSSM becomes unstable and
instabilities develop. These fluctuations grow large, and are expected
to form into Q-balls. The formation of large Q-balls has been studied
only linear theory analytically \cite{KuSh,EnMc1,EnMc2} and numerical
simulations was done in one-dimensional lattices \cite{KuSh}. Both of
them are based on the assumption that the Q-ball configuration is
spherical so that we cannot really tell that the Q-ball configuration
is actually accomplished. Some aspects of the dynamics of AD scalar
and the evolution of the Q-ball were studied in Ref.~\cite{EnMc5}, but
the whole dynamical process was not investigated, which is important
for the investigation of the Q-ball formation.

Actual Q-ball formation is confirmed in our recent work \cite{KK},
where we showed the formation of Q-balls in the gauge-mediated SUSY
breaking scenario using lattice simulations in one, two, and three
dimensions in space. In that scenario, the typical size of Q-balls is
determined by that of the most developed mode of linearized
fluctuations when the amplitude of fluctuations grow as large as that
of the homogeneous mode:
$\langle \delta \phi^2 \rangle \sim \phi^2$. Almost all the initial
charges which the AD condensate carries are absorbed into the formed
Q-balls, leaving only a small fraction in the form of coherently
oscillating AD condensate. Moreover, the actual sizes and the charges
stored within Q-balls depend on the initial charge densities of the AD 
field. We also find that the evolution of the Q-ball crucially depends
on its spatial dimensions, and the stable Q-ball can exist only in the
form of three-dimensional object. 

One may wonder if these results are peculiar to the gauge-mediation
scenario which has a very flat scalar potential for the large field
value. For a very flat scalar potential, larger Q-balls are easily
formed, because the energy of the Q-ball grows $E\sim mQ^{3/4}$
\cite{KuSh}: the larger the charge is, the smaller the energy per unit 
charge is. On the other hand, the Q-ball energy grows linearly in the 
gravity-mediation scenario: $E\sim mQ$ \cite{EnMc2}. Thus, we naively
expect less effective Q-ball formation, particularly for large charge
Q-balls to form. 

In this paper, we show the Q-ball formation in the gravity-mediation
scenario by the use of numerical calculations. We find it very similar
to gauge-mediation version, but some different new features are
revealed. 

In the next section, we see the origin of the fluctuations of the
complex scalar field, and show the instability band. Results from
numerical calculations are shown in Sec.~\ref{numerical}. Here the
charge and the size of Q-balls are found. In Sec.~\ref{baryon},
we will make some comments on the B-ball baryogenesis. We will show
some peculiar phenomena of the Q-ball in the gravity-mediation
scenario, such as the moving Q-balls, and their collisions as a
result. Section \ref{concl} is devoted to our summary and
conclusions. 

\section{Instabilities of Affleck-Dine condensate}
Q-balls with large charge are expected to be formed through
Affleck-Dine mechanism \cite{KuSh}. It is usually considered that
the AD field are rotating homogeneously in its effective potential to
make the baryon numbers. However, if we consider the SUSY-breaking
included potentials, spatial instabilities of the AD field are induced 
by the negative pressure because of its potential being flatter than 
$\phi^2$ \cite{EnMc1,EnMc2,McDonald}. To be concrete, let us
take the following potential \cite{EnMc1,EnMc2}:
\begin{equation}
    \label{pot}
    V(\Phi) = m^2|\Phi|^2\left[ 1+K\log \left( \frac{|\Phi|^2}{M^2}
    \right) \right] 
    - cH^2|\Phi|^2 + \frac{\lambda^2}{M^2}|\Phi|^6,
\end{equation}
where $\Phi$ is a complex scalar field which brings a unit baryon
number, $\lambda$ is a coupling constant of order unity, $H$ is the
Hubble parameter, $c$ is a positive order one constant, $M$ is a large 
mass scale which we take it as $\simeq 2.4\times 10^{18}$ GeV, and the
$K$-term is the one-loop corrections due especially to gauginos, and
the value of $K$ is estimated in the range $-0.01$ to $-0.1$
\cite{EnMc1,EnMc2}. In this potential, the pressure is estimated as
\cite{EnMc1}
\begin{equation}
    P_{\phi} \simeq \frac{K}{2+K}\rho_{\phi} 
             \simeq -\frac{|K|}{2}\rho_{\phi},
\end{equation}
where $\rho_{\phi}$ is the energy density of the scalar field (Here we 
assume that $|K|\ll 1$ so that the first term in Eq.(\ref{pot}) can be 
approximately rewritten in the power-law $\phi^{2+2K}$). Therefore, the
negative value of $K$ is the crucial point for instabilities.

The homogeneous part of the field evolves as
\begin{equation}
    \phi(t) \simeq \left( \frac{a_0}{a(t)} \right)^{3/2} \phi_0,
    \qquad
    \dot{\theta}^2(t) \simeq m^2,
\end{equation}
where we define the field $\Phi$ to be
\begin{equation}
    \Phi(t)=\frac{1}{\sqrt{2}}\phi(t)e^{\theta(t)}.
\end{equation}
Then the equations for the linearized fluctuations can be written as
\begin{eqnarray}
    \label{eom-fl}
    \delta\ddot{\phi} + 3H\delta\dot{\phi}
    - \frac{1}{a^2(t)}\nabla^2\delta\phi 
    - 2\dot{\theta}(t)\phi(t)\delta\dot{\theta}
    - \dot{\theta}^2(t)\delta\phi 
    & & \nonumber \\
    + m^2\left[ 1 + 3K + K\log\left(\frac{\phi^2}{2M^2}\right)\right]
    \delta\phi
    & = & 0, \nonumber \\[5mm]
    \phi(t)\delta\ddot{\theta} 
    + 3H[\dot{\theta}(t)\delta\phi + \phi(t)\delta\dot{\theta}]
    - \frac{\phi(t)}{a^2(t)}\nabla^2\delta\theta
    & & \nonumber \\
    + 2\dot{\phi}(t)\delta\dot{\theta}
    + 2\dot{\theta}(t)\delta\dot{\phi}
    & = & 0,
\end{eqnarray}
We are now going to see the most amplified mode. To this end, we take
the solutions in the form
\begin{equation}
    \delta\phi=\left(\frac{a_0^2}{a^2(t)}\right)^{3/2}\delta\phi_0 
      e^{\alpha(t)+ikx},
    \qquad
    \delta\theta=\delta\theta_0 e^{\alpha(t)+ikx}.
\end{equation}
If $\alpha$ is real and positive, these fluctuations grow
exponentially, and go non-linear to form Q-balls. Putting these forms
into Eqs.(\ref{eom-fl}), we get the following condition for the
non-trivial $\delta\phi_0$ and $\delta\theta_0$,
\begin{equation}
  \label{det}
   \left|
      \begin{array}{cc}
          \ds{F(H)+\ddot{\alpha}
          +\dot{\alpha}^2+\frac{k^2}{a^2}+3m^2K}
          & \ds{-2\dot{\theta}\phi_0\dot{\alpha}} \\
          \ds{2\dot{\theta}\dot{\alpha}}
          & \ds{\left( \ddot{\alpha}+\dot{\alpha}^2+\frac{k^2}{a^2}
            \right) \phi_0}
    \end{array}
    \right| = 0,
\end{equation}
where $F(H)=-\frac{3}{2}\frac{\ddot{a}}{a}-\frac{3}{4}H^2$.

It is appropriate to assume that $H \ll m$ and 
$\ddot{\alpha} \ll \dot{\alpha}$, since the AD field oscillates when
$H \lesssim m$, and the adiabatic production of fluctuations will
occur. Then, Eq.(\ref{det}) will be simplified as
\begin{equation}
    \left( \dot{\alpha}^2+\frac{k^2}{a^2}+3m^2K \right) 
    \left( \dot{\alpha}^2+\frac{k^2}{a^2} \right)
    +4\dot{\theta}^2\dot{\alpha}^2 = 0.
\end{equation}
Since $\dot{\theta}^2\simeq m^2$, for $\dot{\alpha}$ to be real and
positive, we must have 
\begin{equation}
    \frac{k^2}{a^2}\left( \frac{k^2}{a^2}+3m^2K \right) < 0. 
\end{equation}
As we are considering $K$ to be a negative value, an instability band
will exist. This is because the oscillating scalar field in the
potential flatter than $\phi^2$ has negative pressure, which leads to
the instability of the homogeneous field. Thus, the instability band
should be in the range
\begin{equation}
    \label{k-band}
    0 < \frac{k^2}{a^2} < 3m^2|K|.
\end{equation}
We can easily derive that the most amplified mode is the center of the
band: $(k_{max}/a)^2 \simeq 3m^2|K|/2$, and it corresponds exactly to
the Q-ball size which is estimated analytically using the Gaussian
profile of the Q-ball \cite{EnMc2}. We will see shortly that it also
coincides with the size actually observed on the lattices in our
simulations.

\section{Charge and size of Q-balls}
\label{numerical}
In this section, we show the results of the lattice simulations. In
the potential (\ref{pot}), the AD field obeys the equation
\begin{eqnarray}
    \ddot{\Phi}+3H\dot{\Phi}-\frac{1}{a^2}\nabla^2\Phi
    +m^2\Phi\left[1+K+K\log\left(\frac{|\Phi|^2}{M^2}\right)\right]
    & & \nonumber \\
    -cH^2|\Phi|+\frac{3\lambda^2}{M^2}|\Phi|^4\Phi = 0.& &
\end{eqnarray}
Here we have calculated in the matter-dominated universe, so that
$H=2/3t$. In the context of AD mechanism for baryogenesis, the
A-terms, such as $V_{A-term}\sim(A \lambda /M)\phi^4 + h.c.$, should
be added to the potential (\ref{pot}) in order to make the AD field
rotate around in its potential. Instead, we take {\it ad hoc} initial
conditions and neglect A-terms, since they do not affect the later
dynamics of the field crucially. Therefore, the AD field possesses
some initial charge density.

It is more convenient for numerical calculations to take the real and
imaginary decomposition $\Phi=(\phi_1+i\phi_2)/\sqrt{2}$ and rescale 
as follows:
\begin{equation}
\varphi=\frac{\phi}{m}, \qquad h=\frac{H}{m}, \qquad
\tau = mt, \qquad \xi = mx.
\end{equation}
For the initial conditions, we take some large vev in the real axis
and put some angular velocity to the imaginary part. In addition, we
put initial fluctuations very small values $O(10^{-7})$. Thus, they
have the form
\begin{eqnarray}
    \varphi_1(0)=A+\delta A({\bf \xi}), \qquad 
    \varphi'_1(0)=\delta B({\bf \xi}), \nonumber \\
    \varphi_2(0)=\delta C({\bf \xi}), \qquad 
    \varphi'_2(0)=D+\delta D({\bf \xi}),
\end{eqnarray}
where $A$ and $D$ are some constants, independent of the position is
space, $\delta A, \delta B, \delta C,$ and $\delta D$ are ${\bf \xi}$
dependent small random variables, and the prime denotes the derivative 
with respect to $\tau$. Notice that the important features of the
dynamics of the field are not affected by how we take these random
variables, if we do not choose very peculiar distributions. 

\begin{figure}[t!]
\centering
\hspace*{-7mm}
\leavevmode\epsfysize=8cm \epsfbox{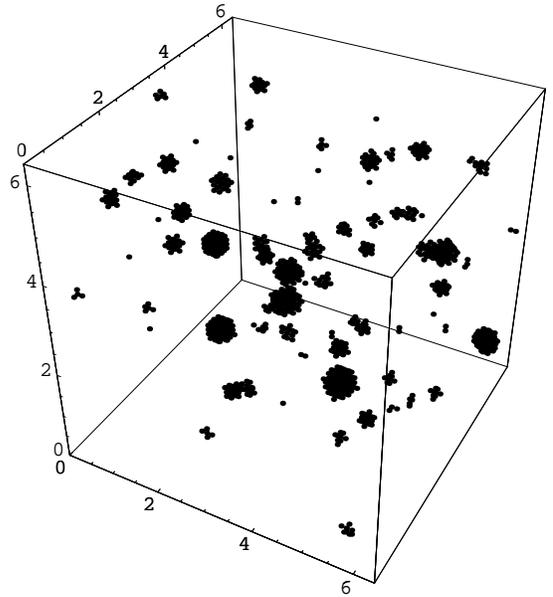}\\[2mm]
\caption[fig-1]{\label{3D-large} 
Configuration of Q-balls on three-dimensional lattice. More than 40
Q-balls are formed, and the largest one has the charge with 
$Q\simeq 5.16\times 10^{16}$}
\end{figure}

\begin{figure}[t!]
\centering
\hspace*{-7mm}
\leavevmode\epsfysize=4cm \epsfbox{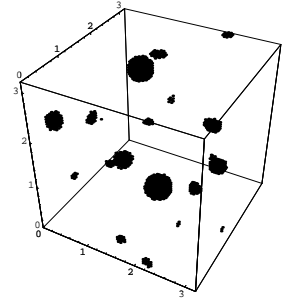}\\[2mm]
\caption[fig-2]{\label{3D-small} 
Configuration of Q-balls on three-dimensional lattice. In each
direction, the box size is half of that in Fig.~\ref{3D-large}. More
than 10 Q-balls are formed, and the largest one has the charge with 
$Q\simeq 1.74\times 10^{16}$}
\end{figure}

We have calculated the dynamics of the AD scalar field for various
parameters, and find that the initially (approximately) homogeneous AD 
field deforms into a lot of clumpy objects. See Figs.~\ref{3D-large}
and \ref{3D-small}. All of them conserve their charge very well, so
they must be Q-balls. (We observed charge loss and exchange between two
Q-balls in some cases. We will discuss them in Sect.~\ref{int-Q}.) The
profile of the Q-ball is a spherically symmetric thick-wall type, and
fits very well to the Gaussian. In these figures, we take 
$\varphi_1(0)=\varphi_2'(0)=2.5\times10^7$ for the initial conditions
on the $64^3$ three-dimensional (3D) lattices with $\Delta\xi=0.1$ and 
$\Delta\xi=0.05$ for the large and small lattice boxes, respectively. 
It seems that there is no box-size effects in these parameters, since
these two figures look the same. They have similar charge
distributions and the Q-ball size is the same, as expected from the
analytical estimate, $R_{phys}\sim |K|^{-1/2}m^{-1}$. Actually, the
numbers of Q-balls with the charge larger than $10^{15}$ are 7 and 2
in the large and small box, respectively.

Comparing to those Q-balls which appear in the gauge-mediated SUSY
breaking scenario, the size of the Q-ball is much smaller for the same 
charge, and most of the Q-balls has the same order of size. This is
because $R_{phys} \sim |K|^{-1/2}m^{-1}$ for the gravity-mediation,
which does not depend on the charge $Q$, while 
$R_{phys} \sim m^{-1}Q^{1/4}$ for the gauge-mediation. We thus observe 
large-charged Q-balls with relatively small size.

\begin{figure}[t!]
\centering
\hspace*{-7mm}
\leavevmode\epsfysize=7cm \epsfbox{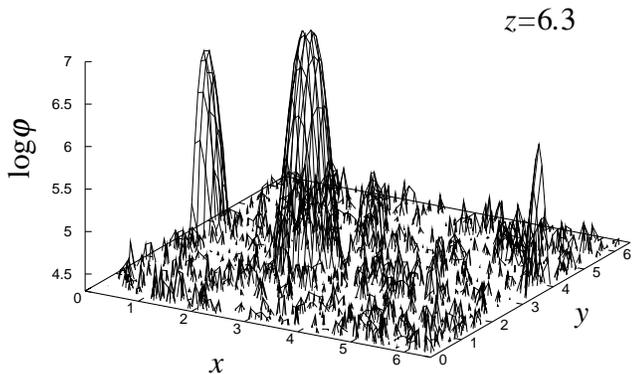}\\[2mm]
\caption[fig-3]{\label{relic3D} 
Amplitude of the AD field after formation of Q-balls. This
configuration is the slice at $z=6.3$. The amplitude of relic field
outside the Q-balls is two or three orders smaller than that of the
center of the Q-balls.}
\end{figure}

\begin{figure}[t!]
\centering
\hspace*{-7mm}
\leavevmode\epsfysize=6cm \epsfbox{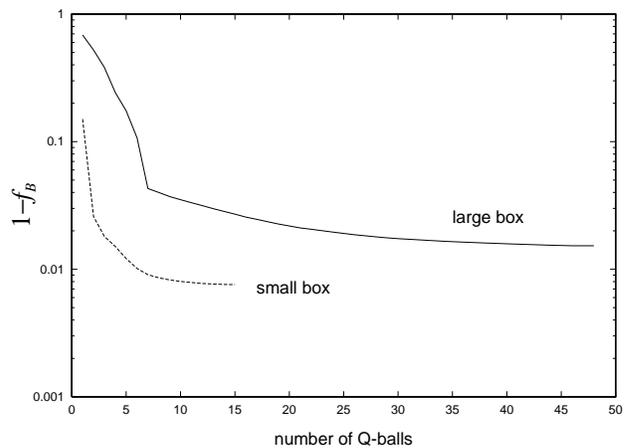}\\[2mm]
\caption[fig-3]{\label{totQ} 
Fraction of the charge outside the Q-balls.The solid and dotted lines
denote the results from the simulations shown in Figs.~\ref{3D-large}
and \ref{3D-small}, respectively.}
\end{figure}

As in the case of the gauge-mediation scenario \cite{KK}, we observe
almost all the charge which initially AD condensate has absorbed into
Q-balls, and the amplitude of the relic AD field is highly damped. 
This means that the fraction of the charge outside Q-balls is very
small. Figure~\ref{relic3D} shows the amplitude of the AD field of
the slice at $z=6.3$ in the larger box for another realization of
simulations. Notice that there is relic field outside Q-balls, but the 
fluctuations are rather large, and we may not be able to consider it
as a homogeneous condensate.  In particular case of 
Figs.~\ref{3D-large} and \ref{3D-small}, Q-balls carry more than
$97\%$ and $99\%$ of the total charge, respectively. 
In Fig.~\ref{totQ}, the fraction of the charge outside the Q-balls is
shown as a function of the number of Q-balls which we take into
account. In the larger box simulation, only seven of the largest
Q-balls hold more than $95\%$ of the total charge. On the other hand,
more than $97\%$ is stored in only two of the largest Q-balls in the
small box one. Notice that the dotted line (small box) is below the
solid line (large box), because the resolution is twice as good in the
former simulation: the lower bound is determined by the resolution of
each simulations.

\begin{figure}[t!]
\centering
\hspace*{-7mm}
\leavevmode\epsfysize=6cm \epsfbox{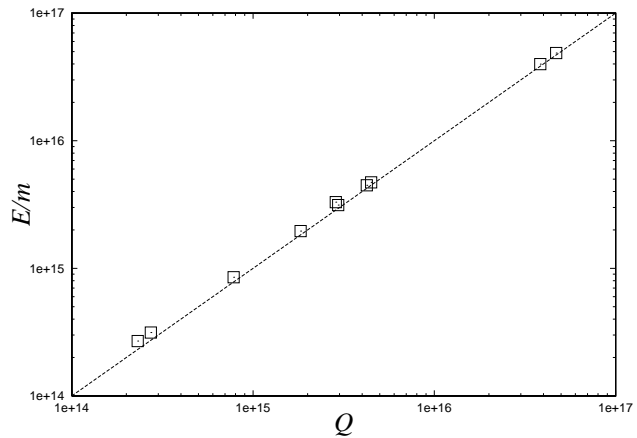}\\[2mm]
\caption[fig-3]{\label{e-q} 
Dependence of the energy of the Q-ball on its charge calculated on
three-dimensional lattices. This confirms the analytical estimate:
$E\simeq mQ$ (the dotted line).}
\end{figure}

Analytically, some features of the Q-ball in gravity-mediation is
known \cite{EnMc2}. For example,
\begin{equation}
    \label{relation}
    E \sim mQ, \quad, R_{phys}\sim |K|^{-1/2}m^{-1},
    \quad \omega \sim m, \quad {\rm etc.}
\end{equation}
They are all confirmed numerically. One example is shown in
Fig.~\ref{e-q}. This confirms the first relation of
Eq.(\ref{relation}), which implies that the energy per unit charge is 
constant of $O(m)$.

It is the best way to investigate the dynamics of Q-ball formation on
{\it three}-dimensional lattices, but it is practically difficult to
do, since we need somewhat high resolution, and many runs for various
parameters to look at. Thus, we also calculate on one and 
two-dimensional lattices for more rigorous quantitative
analysis. Therefore, we must know the evolution of Q-balls after their 
formation. We follow the similar discussion we made for Q-balls in
the gauge-mediation scenario \cite{KK}. Since a Q-ball configuration
is the energy minimum with some fixed charge $Q$, $Q$ is constant with 
respect to time, so
\begin{equation}
    Q = a^3 Q_D \sim a^3 R^D \tilde{q} \sim {\rm const},
\end{equation}
where $Q_D$ is the charge in $D$ dimension, and 
$\tilde{q}=\phi_1\dot{\phi}_2-\dot{\phi}_1\phi_2$ is the charge
density. If we assume the form of a Q-ball as
\begin{equation}
  \phi({\bf x},t)=\phi({\bf x})\exp({i\omega t}),
\end{equation}
the energy of a Q-ball can be calculated as
\begin{eqnarray}
  E & = & \int d^3x \left[ \frac{1}{2}(\nabla\phi)^2 + V(\phi)
        -\frac{1}{2}\omega^2\phi^2 \right] +\omega Q, \nonumber \\
    & = & \int d^3x [ E_{grad} + V_1 + V_2 ] + \omega Q,
\end{eqnarray}
where 
\begin{eqnarray}
  E_{grad} & \sim & \frac{\phi^2}{a^2R^2}, \nonumber \\
  V_1      & \sim & m^2M^{2|K|}\phi^{2-2|K|}, \nonumber \\
  V_2      & \sim & \omega^2\phi^2.
\end{eqnarray}
Here we assume that the logarithmic term of the first term in the
potential (\ref{pot}) is small compared to the unity, so that we can
approximate it in the power-law form.

When the energy take the minimum value, the equipartition is achieved: 
$E_{grad} \sim V_1$ and $E_{grad} \sim V_2$. From these equations
addition to the charge conservation, we obtain the following
evolutions:
\begin{eqnarray}
  R       & \propto & a^{-(1+2|K|)/[1+(D-1)|K|]}, \nonumber \\
  \phi    & \propto & a^{-(3-D)/[1+(D-1)|K|]}, \nonumber \\
  \omega  & \propto & a^{(3-D)|K|/[1+(D-1)|K|]},
\end{eqnarray}
which we observed approximately the same features numerically. For
$D=3$, we get very natural relations: $R_{phys}=Ra\sim$ const., 
$\omega\sim$ const., and $\phi\sim$ const. Although $\phi$ decreases
as time goes on for $D=1$ and $2$, $R$ and $\omega$ is almost
constant, since $|K|\ll 1$. This feature is different from that in the 
gauge mediation scenario, and is good for long simulations because 
low-dimensional Q-balls do not shrink the size so much.

Now we will see that the size of the Q-ball is determined by the most
amplified mode. Comparing to the actual sizes observed on lattices, we 
also calculated numerically for linearized fluctuations. Although we
decomposed the complex field in radial and phase direction in the
previous section, it is more convenient to decompose it into real and
imaginary part for numerical simulations. We thus integrated the
following mode equations in dimensionless variables:
\begin{eqnarray}
& & \delta\varphi_i''+3h\varphi_i'
    + \left[ \frac{k^2}{a^2} + 1+K
        +K\log\left( 
       \frac{\tilde{m}^2(\varphi_1^2+\varphi_2^2)}{2} \right) 
      \right.  \nonumber \\ 
& & \left. \hspace{25mm}
       +2K\frac{\varphi_i^2}{\varphi_1^2+\varphi_2^2} 
       -ch^2  \right. \nonumber \\ 
& & \left. \hspace{25mm}
       +\frac{3}{4}\lambda^2\tilde{m}^2(5\varphi_i^2+\varphi_j)
           (\varphi_1^2+\varphi_2^2) \right] \delta\varphi_i
    \nonumber \\ 
& & \hspace{40mm}
    + 2K\frac{\varphi_1\varphi_2}{\varphi_1^2+\varphi_2^2}
    \delta\varphi_j =0, 
\end{eqnarray}
where $(i,j)=(1,2)$, $(2,1)$, and $\tilde{m}=m/M$.

\begin{figure}[t!]
\centering
\hspace*{-7mm}
\leavevmode\epsfysize=8cm \epsfbox{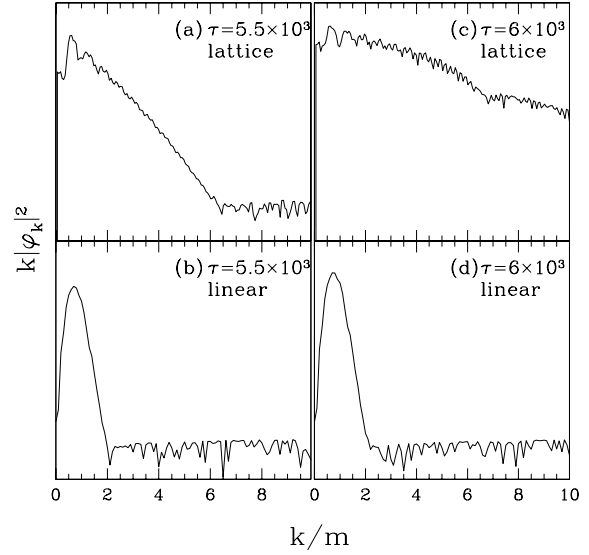}\\[2mm]
\caption[fig-3]{\label{spect} 
Power spectra of fluctuations of AD scalar field
($k^D|\delta\varphi_k|^2$, $D=1$) when the amplitude of fluctuations
becomes as large as that of the homogeneous mode: $\langle \delta
\varphi^2 \rangle \sim \varphi^2$. The top panels (a) and (c) show the
full fluctuations calculated on one dimensional lattices, while the
bottom panels (b) and (d) show the linearized fluctuations without
mode mixing.}
\end{figure}

Figure \ref{spect} shows the power spectrum calculated from a lattice
simulation and the above linearized equations at $\tau=5.5\times 10^3$ 
and $\tau=6\times 10^3$. We take the lattices with lattice size
$N=1024$ and lattice spacing $\Delta\xi=0.1$ in one dimension here,
because we need high resolution data to make the power spectrum smooth
for lower $k$. These two times are just before and after the
fluctuations are fully developed: 
$\langle\delta\varphi^2\rangle\sim\varphi^2$. For linearized
fluctuations, the instability band is almost the same as 
Eq.(\ref{k-band}). For example, the upper bound is estimated by 
$k/m=\sqrt{3}a(\tau)|K|^{1/2}\approx 2.5$ for $|K|=0.01$ and
$\tau=5.5\times10^3$. See panel (b). Even before the full
development of fluctuations (panel (a)), rescattering effects
kick the lower mode to higher, and the spectrum gets a little broader
\cite{KhTk}. Needless to say, the spectrum becomes extremely broad
and smooth after  fluctuations are fully developed (panel (c)). At
any times, however, the peeks are at the same points for both lattices
and linearized cases, and correspond to the typical size of Q-balls
actually observed on the lattices. Therefore, we can conclude that the
size of the Q-ball is determined by the most amplified mode of the
linearized fluctuations when they are fully developed. For the case of 
Fig.~\ref{spect}, the typical size is $k_{max}\sim 0.5$, which
implies $R_{phys} \sim a(\tau_f)/k_{max} \sim 28.9$, where 
$\tau_f \simeq 5.5\times10^3$ is the formation time. This value
exactly coincides with the sizes of Q-balls observed on 
three-dimensional lattices. Actually, they are (a few)$\times 10$ in
the dimensionless units.

\begin{figure}[t!]
\centering
\hspace*{-7mm}
\leavevmode\epsfysize=6cm \epsfbox{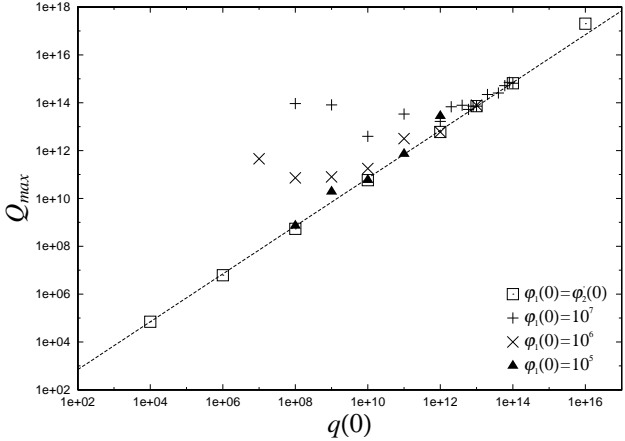}\\[2mm]
\caption[fig-4]{\label{q-dep} 
Dependence of charges on the initial charge density
$q(0)=\varphi_1(0)\varphi_2'(0)$ carried by the AD condensate on 
one-dimensional lattices. Open squares denote the case 
$\varphi_1(0)=\varphi_2'(0)$, ``pluses,''crosses, and solid triangles
denote the dependence on $\varphi_2'(0)$ with $\varphi_1(0)$ fixed at 
$10^7$, $10^6$, and $10^5$, respectively.}
\end{figure}

The actual values of the charge depend on the values of the charge
density which AD field initially possesses. Since initial charge
density is written as $q(0)=\varphi_1(0)\varphi_2'(0)$ for our initial 
conditions, we must check the dependence on both initial amplitude
$\varphi_1(0)$ and angular velocity $\varphi_2'(0)$ of AD
field. Results are shown in Fig.~\ref{q-dep}. Here we plot the largest 
charge $Q_{max}$ against the initial AD charge density $q(0)$. We
investigate two situations. The first one is changing both equally
while fixing the relation $\varphi_1(0)=\varphi_2'(0)$, which is shown
by open squares in the figure. This corresponds to the ``maximum
charged'' Q-balls in terms of Ref.~\cite{EnMc5}. We can fit all of
these on the straight line (dotted line), 
$Q_{max}\approx 7\times q(0)$, and the Q-ball charge depends linearly
on the initial charge density. 
 
The second situation is the dependence on the angular velocity 
$\varphi_2'(0)$ while $\varphi_1(0)$ is fixed. We calculate for three
different value of $\varphi_1(0)$: $10^7$, $10^6$, and $10^5$. In all
cases, linear dependence is still preserved when the ratio of 
$\varphi_1(0)$ and $\varphi_2'(0)$ is within two orders of magnitude.
However, if $\varphi_2'(0)$ becomes smaller, the maximum Q-ball charge
does not depend on the initial charge density. This is due to the
creation of the negative-charged Q-balls. The charge is determined
only by $\varphi_1(0)$. 

\begin{figure}[t!]
\centering
\hspace*{-7mm}
\leavevmode\epsfysize=6cm \epsfbox{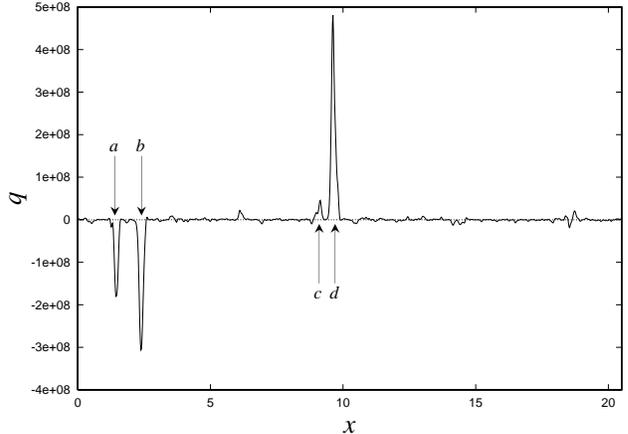}\\[2mm]
\caption[fig-2]{\label{negative} 
Configuration of positive and negative Q-balls on one-dimensional
lattice. Here we take $\varphi_1(0)=10^7$ and $\varphi_2'(0)=10^2$.
The four largest Q-balls have the charges (a) $-2.9\times10^{13}$, 
(b) $-5.6\times10^{13}$, (c) $6.8\times10^{12}$, and 
(d) $8.1\times 10^{13}$.}
\end{figure}

Negative charge Q-balls are formed when the (initial) angular velocity 
is rather small. Figure~\ref{negative} shows an example. In this case, 
we see the largest Q-ball with positive charge, two large negative
charge Q-balls, and one Q-ball with positive charge an order of
magnitude smaller for four largest ones. Similar situations occur in
the gauge mediation scenario \cite{KK}, but the critical value of the
ratio $\varphi_2'(0)/\varphi_1(0)$ for the negative charge Q-ball
formation is larger in the gravity mediation scenario. This is because
the angular motion of the AD condensate is more circular and stable,
and the produced Q-ball size is larger in the flatter potential, so
that it is more difficult to reverse the angular velocity of the field 
within that size.

\begin{figure}[t!]
\centering
\hspace*{-7mm}
\leavevmode\epsfysize=6cm \epsfbox{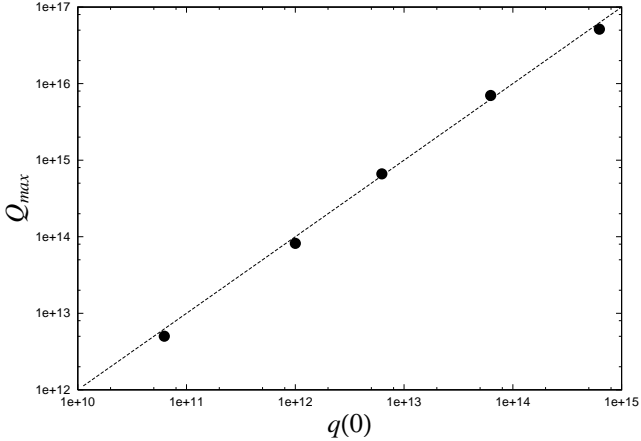}\\[2mm]
\caption[fig-4]{\label{q-dep3D} 
Dependence of charges on the initial charge density
$q(0)=\varphi_1(0)\varphi_2'(0)$ carried by the AD condensate on 
three-dimensional lattices. }
\end{figure}

In the actual situation, the AD field takes a very large vev before
it rolls down to the origin of its potential, and the vev is
determined by equating second and third terms in the potential
(\ref{pot}):
\begin{equation}
    \phi \sim \sqrt{\frac{HM}{\lambda}}. 
\end{equation}
The AD field begins to roll down when $H\sim m$, so its amplitude is
$\varphi\sim (\lambda\tilde{m})^{-1/2}\simeq 2.4\times 10^7$ in the
dimensionless parameters, where $\tilde{m}=m/M$. At the same time, the 
AD field begins rotation because of the A-term of the form, 
$V_{A-term}\sim(\lambda m/M)\phi^4 + h.c.$ If we assume that the
initial angular velocity is the same order as the initial amplitude in 
the dimensionless units, we get the initial charge density as
$q(0) = \varphi_1(0)\varphi_2'(0) \sim 6\times10^{14}$. We expect the
linear dependence between the initial charge density of the AD
condensate and the produced largest Q-ball on three-dimensional
lattices, as $Q_{max}\simeq q(0)\times 10^2$. This is shown in 
Fig.~\ref{q-dep3D}, where we take such initial conditions as the
linear dependence is expected to hold, i.e., 
$\varphi_1(0)\sim \varphi_2'(0)$. Using this relation, we can estimate
the maximum charge of the actually expected Q-balls is 
$Q_{max}\sim 6\times 10^{16}$. For the B-ball baryogenesis to work,
the charge should exceed $10^{20}$ \cite{EnMc2}. Therefore, it may be
a little difficult to reach this value in the parameters in the
model. However, if we take $\lambda^2\phi^{10}/M^6$ instead of 
$\lambda^2\phi^6/M^2$ in the potential, as appears in the $u^cd^cd^c$
flat direction \cite{EnMc1,EnMc2}, the initial vacuum expectation
value (vev) of the AD field is estimated as 
$\varphi \sim (\lambda\tilde{m}^3)^{-1/4}\simeq7\times10^{10}$. In
this case, the initial AD charge density becomes $\sim5\times10^{21}$,
and it implies that the maximum Q-ball charge reaches as large as
$\sim 5\times 10^{23}$. Thus, we get enough value of the charge for
B-ball baryogenesis. 

\section{B-ball baryogenesis and its restrictions to the particle
physics} 
\label{baryon}
As is known, baryon number and the amount of the dark matter can be
directly related in the B-ball baryogenesis in the gravity-mediated
SUSY breaking scenario \cite{EnMc2}. To this end, it is important to
estimate how much charges are stored in the form of the Q-ball. In
some cases, the fraction of the Q-ball charge may restrict the mass of
the LSP, and vice versa \cite{EnMc2,EnMc3}. We have calculated for
various initial conditions on one-, two-, and three-dimensional
lattices, and find that almost all the charges are absorbed into
Q-balls. This fact 
is also true when we take other values for parameters in the
potential. In particular, we investigate for the fraction of Q-ball
charge, changing $K$ from $-0.01$ to $-0.1$. It was  done by other
method in Refs.~\cite{EnMc5}, and they concluded that when the
absolute value $|K|$ was larger, the less the fraction. However, our
results differ from theirs. We collect them in Tables~\ref{K-1D} and
\ref{K-3D}. The former is the results from one-dimensional lattices
with the box size $N\Delta\xi=1024\times0.02=20.48$. The latter table
shows the results calculated in three dimensions. In this case, the
box size is $N\Delta\xi=64\times 0.1=6.4$. As can be seen, the
fraction of the sum of charge of Q-balls to the total charge has no
dependence on the value of $K$. Moreover, neither does it depend upon
the ratio of $\varphi_1(0)$ and $\varphi_2'(0)$. All of them lead to a
conclusion that almost all the charges are stored in Q-balls: that is,
$f_B \approx 1$.

\begin{table}
\caption{Fraction of the charge stored in Q-balls for various values
of $K$ and $\varphi_2'(0)/\varphi_1(0)$ on one-dimensional lattices.}
\label{K-1D}
\renewcommand{\arraystretch}{1.2}
\begin{tabular}{clll}
$\varphi_2'$  $\setminus$ K & $-0.01$  & $-0.05$  & $-0.1$   \\ \hline 
$1.0\times10^7$  & 95.2\% & 98.6\% & 93.0\% \\
$8.0\times10^6$  & 97.3   & 98.2   & 98.9   \\
$6.0\times10^6$  & 98.0   & 99.9   & 99.7   \\
$4.0\times10^6$  & 99.1   & 97.9   & 98.6   \\
$2.0\times10^6$  & 99.0   & 97.6   & 98.3   \\
$1.0\times10^6$  & 91.5   & 97.5   & 99.6   \\
$8.0\times10^5$  & 97.6   & 95.5   & 97.0   \\
$6.0\times10^5$  & 96.1   & 97.4   & 97.9   \\
$4.0\times10^5$  & 99.4   & 95.2   & 99.7   \\
\end{tabular}
\end{table}

\begin{table}
\caption{Fraction of the charge stored in Q-balls for various values
of $K$ and $\varphi_2'(0)/\varphi_1(0)$ on three-dimensional
lattices.}
\label{K-3D}
\renewcommand{\arraystretch}{1.2}
\begin{tabular}{rlll}
$\varphi_2'$  $\setminus$ K & $-0.01$  & $-0.05$  & $-0.1$ \\ \hline
$2.5\times10^7$  & 98.7\% & 99.7\% & 99.1\% \\
$2.5\times10^6$  & 98.1   & 99.4   & 99.5   \\
$2.5\times10^5$  & 98.4   & 99.8   & 99.2   \\
\end{tabular}
\end{table}

Following the argument of Refs.~\cite{EnMc2,EnMc3}, the number density
of the baryon to that of the dark matter ratio can be written in terms
of density parameters as 
\begin{equation}
    \frac{n_b}{n_{DM}}=
    \frac{\Omega_b}{\Omega_{DM}}\frac{m_{DM}}{m_N},
\end{equation}
where $m_N\simeq 1$ GeV is the nucleon mass. In the B-ball baryogenesis 
of the gravity-mediation scenario, B-balls decay into baryons and
LSP neutralinos, so that the relation between the number density of
baryon and dark matter is $n_{DM}=N_B f_B n_b$, where $N_B$ is the
number of neutralinos into which the AD field with a unit charge
decays, and it is usually $\gtrsim 3$. Here we assume no later
annihilation of neutralinos. Using the conservative constraint on the
amount of the baryon number from the Big-Bang nucleosynthesis, 
$0.004 \lesssim \Omega_b h^2 \lesssim 0.023$ \cite{Olive}, we get a
stringent constraint on the neutralino mass  
\begin{equation}
    7.1 {\rm GeV} \lesssim m_{\chi}\left( \frac{N_B}{3}\right)
    \left( \frac{\Omega_{DM}h^2}{0.49}\right)^{-1} f_B
    \lesssim 40.8 {\rm GeV}.
\end{equation}
This bound is marginally consistent with $f_B\approx 1$ and the
accelerator experiment bounds such as $M_{\chi}\gtrsim 24.2$ GeV
\cite{PDG}. Note that the constraint becomes more severe if
$\Omega_{DM}$ is smaller than 1 as in the case, for example, that
considerable fraction of the total energy density is stored in the
form of the cosmological constant \cite{EnMc1,EnMc2}. In this case,
the annihilation of neutralinos must be taken place.

\section{moving Q-balls, their interactions, and Breather-like
soliton} 
\label{int-Q}
\begin{figure}[t!]
\centering
\hspace*{-7mm}
\leavevmode\epsfysize=8cm \epsfbox{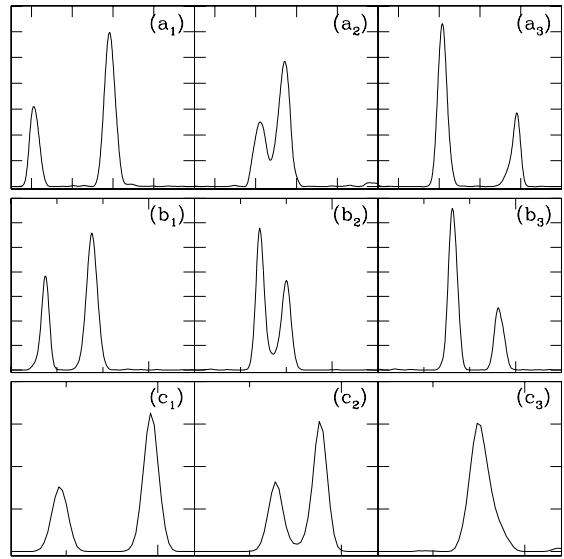}\\[2mm]
\caption[fig-4]{\label{move} 
Configurations of Q-balls for (a) passing through, (b) exchanging part 
of charges, and (c) merging.}
\end{figure}

As the consequence that the size of Q-balls is relatively small in the
gravity-mediated SUSY breaking scenario, in a fixed volume, the
coherent AD field breaks into larger numbers of Q-balls than in the
gauge mediation scenario. Therefore, Q-balls can have somewhat large
peculiar velocities, as opposed to Q-balls in gauge-mediation
scenario. Actually, we observed moving Q-balls on the lattices in one,
two, and three dimensions, but, unfortunately, Q-ball collisions
(interactions) are observed only on one-dimensional lattices. This is
not a surprise, since the impact parameter is small for small size
Q-balls in two or three dimensions. On the other hand, in one
dimension, Q-balls must collide if they have enough (initial)
velocities. We see the following three patterns for the
interactions:(a) passing through, (b) exchanging part of charges, and
(c) merging. They are expressed symbolically as 
\begin{mathletters}
\label{int}
\begin{eqnarray}
    \label{pass}
    A + B & \longrightarrow & B+A, \\
    \label{exchange}
    A + B & \longrightarrow & B'+A',\\
    \label{merge}
    A + B & \longrightarrow & C.
\end{eqnarray}
\end{mathletters}
These situations are plotted in Fig.~\ref{move}. For the top three
panels, they show the type (a), and two Q-balls with charges 
$4.0\times10^{15}$ and $1.8\times10^{15}$ are approaching, get
together with the charge $5.8\times10^{15}$, and finally pass through
each other without changing their own charges. For the middle three
panels, they represents the type (b). They exchange about 10\% of
their charges. In the bottom three panels, we show the merging
process.

\begin{figure}[t!]
\centering
\hspace*{-7mm}
\leavevmode\epsfysize=8cm \epsfbox{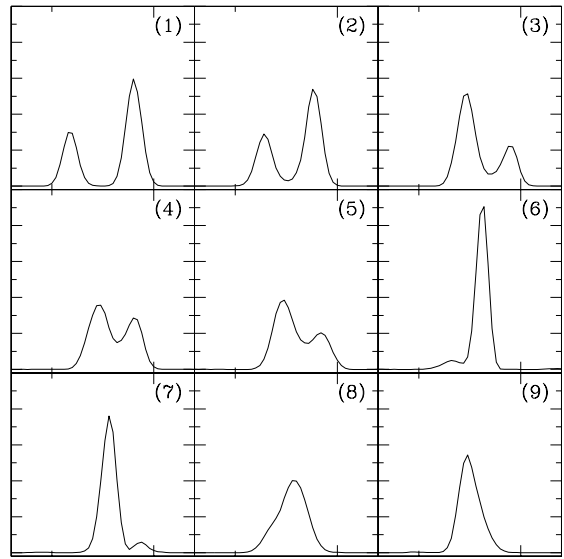}\\[2mm]
\caption[fig-4]{\label{breather} 
Configurations of merging Q-balls on one-dimensional lattices. Each of 
the panels show the time snapshots at from (1) $\tau=4.375\times10^4$
to (9) $\tau=4.775\times10^4$ with the interval
$\Delta\tau=0.05\times10^4$.}
\end{figure}

Qualitatively, these processes can be divided by the relative velocity
of two colliding  Q-balls. If the relative velocity is large, they
pass through each other without any (or negligible) charge exchange. 
When the velocity are smaller, two Q-balls exchange part of their
charges. When the velocity is still slower, they merge into one, and
it vibrates for a while. It can be a breather-like soliton, and an
example is shown is Fig.~\ref{breather}. It repeats the double peaks
and the single peak profiles just after the collision until it becomes
stable state. During this process, we observed the decay of the charge
by emitting very small Q-balls. For this particular example, about
$7\%$ of its charge is lost until it finally becomes stable and 
conserves its charge from that time on. The decrease of charge can be
explained also by the emission of scalar waves, but we cannot
distinguish them in the resolution of our simulations. In addition to
the merging process (c), we see a few inverse processes: the breaking
into two. These three processes (a), (b), and (c) are very similar to
the results of Ref.~\cite{AKPF}, where the collision of
non-topological solitons for other type is studied numerically on 
two-dimensional lattices. Although we do not have a chance to see any
collision in two or three dimensions, their properties may be very
similar if it happens to occur. 

\section{Conclusions}
\label{concl}
We have calculated the full non-linear dynamics of the complex scalar
field, which represents some flat direction carrying the baryonic
charge in MSSM, in the context of the gravity-mediated SUSY breaking
scenario. Since the scalar potential in this model is flatter than
$\phi^2$, we have found that fluctuations develop and go non-linear to
form non-topological solitons, Q-balls. As in the gauge-mediation
scenario \cite{KK}, the size of a Q-ball is determined by the most
amplified mode, but this mode is completely determined by the model
parameters $m$ and $K$, and the size does not depend on the charge
$Q$. On the other hand, the charge of Q-balls depends on the initial
charge density of the Affleck-Dine field, and its dependence is
linear. Therefore, large-charged Q-balls with relatively small size
are formed in this scenario. 

Once Q-balls are formed, almost all the charges are absorbed into them
in all the simulations we made, and only a tiny fraction of the charge
is carried by the relic AD field, but its amplitude is very small and
fluctuates so that it may not be possible to regard it as a
condensate. This leads to some interesting results. We can restrict
the scenario of the baryogenesis, which has a direct relation to the
amount of the dark matter, or the parameter in MSSM, such as the
neutralino mass, can be constrained.

We have also observed moving Q-balls, which is peculiar to the
gravity-mediation scenario. In this case, larger numbers of Q-balls
are formed in a fixed box size because of the relatively small Q-ball
size, so the peculiar velocities are larger than those in the
gauge-mediation scenario. As a consequence, there are collisions of
Q-balls. The probability of collision crucially depends on the
spatial dimensionality, and we have not found any collision in
two or three dimensions. We thus expect the probability to be small in
an actual situations. However, very interesting phenomena will occur,
if collisions happen to take place. They are the charge exchange and
merging to be large charge Q-balls. If the charge of a Q-ball becomes
larger, it will be more difficult to evaporate or to be dissociated.

\section*{Acknowledgement}
M.K. is supported in part by the Grant-in-Aid, Priority
Area ``Supersymmetry and Unified Theory of Elementary
Particles''($\#707$).

\end{document}